\documentclass[ aps, showpacs, showkeys, nofootinbib, floatfix, superscriptaddress]{revtex4}

\usepackage{amsfonts}

\usepackage{amssymb}
\usepackage{amsmath}
\usepackage{graphicx}

\begin{document}

\title{Acceleration of the cosmic expansion induced by symmetry breaking}

 \author{G. Y. Chee}
 \email{qgy8475@sina.com}
 \affiliation{College of Physics and Electronics, Liaoning Normal University, Dalian,\\
 116029, China}

\begin{abstract}

It is proved that in order to obtain a model of the accelerated cosmic
expansion the thing one only need to do is to add a perturbation term to the
Einstein-Hilbert Lagrangian. This term leads to some symmetry breaking terms
in the fields equation, which makes the cosmic expansion accelerating. A
vacuum de Sitter solution is obtained. A new explanation of the acceleration
of the cosmic expansion is presented. In this model the changing of the
expansion from decelerating to accelerating is an intrinsic property of the
universe without need of an exotic dark energy. The acceleration of the
cosmic expansion is induced by the symmetry breaking perturbation of the
gravitational energy. The cosmological constant problem, the coincidence
problem and the problem of phantom divide line crossing are naturally
solved. The results of the model are roughly consistent with the
observations.

\end{abstract}



\maketitle

\section{$\text{Introduction}$}

There is, by now, convincing evidence for new physics associated with the
gravitational interaction on cosmological distance scales \cite{1}. It is
fashionable to attribute these phenomena to esoteric sources of
energy-momentum \cite{2}, but it is conceivable that gravity itself should
be modified by adding terms to the action for General Relativity \cite{3}.
Very recently, there have been proposals for constructing generalizations of
teleparallel gravity \cite{4} which followed the spirit of modified gravity
\cite{3} as a generalization of general relativity. The interest in these
theories was aroused by the claim that their dynamics differ from those of
general relativity but their equations are still second order in derivatives
and, therefore, they might be able to account for the accelerated expansion
of the universe and remain free of pathologies. It has been shown, however,
that these theories are not locally Lorentz invariant \cite{5}. Gravity is
usually considered to be irrelevant as far as the physics of elementary
particles is concerned and, in particular, in the context of the spontaneous
symmetry breaking mechanism. For many years only fewer works \cite{6} have
been done in this direction and no work suggests a relation between symmetry
breaking and cosmological phenomena. The aim of this letter is to show that
the acceleration of the cosmic expansion can be produced by a symmetry
breaking. By adding a perturbation term to the Einstein-Hilbert Lagrangian a
new cosmological model is obtained. In this model the field equation include
some symmetry breaking terms which play the role of the so-called ''dark
energy'' and make the cosmic expansion accelerate. A vacuum de Sitter
solution is obtained. The cosmological constant problem, the coincidence
problem and the problem of phantom divide line crossing are naturally
solved. The coefficient of the perturbation term is found to be a new
cosmological parameter describing the evolution of the universe.

\section{$\text{Field equations}$}

We start from the action
\begin{equation}
S=\frac 1{2\kappa ^2}\int \sqrt{-g}\left( R+\alpha \Gamma ^2+{\cal L}%
_m\right) d\Omega ,
\end{equation}
where $\kappa ^2=8\pi G_N$ with the bare gravitational constant $G_N$,
\begin{equation}
\Gamma =\left( \left\{ _\sigma {}^\mu {}_\nu \right\} \left\{ _\rho {}^\nu
{}_\mu \right\} -\left\{ _\mu {}^\nu {}_\nu \right\} \left\{ _\sigma {}^\mu
{}_\rho \right\} \right) g^{\rho \sigma },
\end{equation}
with the Christoffel symbol $\left\{ _\nu {}^\mu {}_\sigma \right\} $, and $%
\alpha $ is a new parameter describing the evolution of the universe. We
will see that $\alpha $ is very small and then the term $\alpha \Gamma ^2$
can be considered as a perturbation term. The variational principle yields
the field equations for the metric $g_{\mu \nu }$:

\begin{equation}
\left( 1+2\alpha \Gamma \right) \left( R_{\rho \sigma }-\frac 12Rg_{\rho
\sigma }\right) +\frac \alpha 2g_{\rho \sigma }\Gamma ^2-2\alpha \Gamma
_{,\mu }\frac{\partial \Gamma }{\partial g^{\rho \sigma }{}_{,\mu }}=\kappa
^2T_{\rho \sigma }.
\end{equation}
where $R_{\rho \sigma }$ is the Ricci curvature tensor of $\left\{ _\sigma
{}^\mu {}_\nu \right\} $, $R=g^{\rho \sigma }R_{\rho \sigma }$, and
\[
T_{\rho \sigma }=-2\frac{\delta {\cal L}_m}{\delta g^{\rho \sigma }}
\]
is the energy-momentum of the matter fields. These equations can be
re-arranged in the Einstein-like form
\begin{equation}
R_{\mu \nu }-\frac 12Rg_{\mu \nu }=-\frac{\alpha g_{\mu \nu }\Gamma ^2}{%
2\left( 1+2\alpha \Gamma \right) }+\frac{2\alpha \Gamma _{,\lambda }}{\left(
1+2\alpha \Gamma \right) }\frac{\partial \Gamma }{\partial g^{\mu \nu
}{}_{,\lambda }}+\frac{\kappa ^2}{\left( 1+2\alpha \Gamma \right) }T_{\mu
\nu }.
\end{equation}
The first and the second term on the right are not covariant and represent
the ''dark'' energy-momentum which has the same origin with the
energy-momentum of the gravitational field. We will see that it is this
perturbational gravitational energy-momentum that induces the acceleration
of the cosmic expansion.

\section{$\text{Cosmological model}$}

We now investigate the cosmological dynamics for the models of this theory.
We consider a flat Friedmann-Lemaitre-Robertson-Walker spacetime with the
metric
\begin{equation}
g_{\mu \nu }=\text{diag}\left( -1,a\left( t\right) ^2,a\left( t\right)
^2,a\left( t\right) ^2\right) ,
\end{equation}
where $a(t)$ is a scale factor. For the matter Lagrangian ${\cal L}_m$ in
Eq. (1) we take into account non-relativistic matter and radiation. The
non-vanishing components of the Christoffel symbol are

\begin{eqnarray}
\left\{ _0{}^0{}_0\right\}  &=&0,\left\{ _0{}^0{}_i\right\} =\left\{
_i{}^0{}_0\right\} =0,\left\{ _i{}^0{}_j\right\} =a\stackrel{\cdot }{a}%
\delta _{ij},  \nonumber \\
\left\{ _0{}^i{}_0\right\}  &=&0,\left\{ _j{}^i{}_0\right\} =\left\{
_0{}^i{}_j\right\} =\frac{\stackrel{\cdot }{a}}a\delta _j^i,\left\{
_j{}^i{}_k\right\} =0,i,j,k,...=1,2,3.
\end{eqnarray}
and
\begin{equation}
\Gamma =\left( \left\{ _\sigma {}^\mu {}_\nu \right\} \left\{ _\rho {}^\nu
{}_\mu \right\} -\left\{ _\mu {}^\nu {}_\nu \right\} \left\{ _\sigma {}^\mu
{}_\rho \right\} \right) g^{\rho \sigma }=-6\frac{\stackrel{\cdot }{a}^2}{a^2%
}=-6H^2,
\end{equation}
where $H\equiv \stackrel{\cdot }{a}/a$ is the Hubble parameter and a dot
represents a derivative with respect to cosmic time $t$. The the field
equations (4) take the forms :
\begin{equation}
3H^2=\kappa ^2\left( \rho +\rho _{\text{de}}\right) ,
\end{equation}
and
\begin{equation}
-2\stackrel{\cdot }{H}-3H^2=\kappa ^2\left( p+p_{\text{de}}\right) ,
\end{equation}
or
\begin{equation}
\frac{\stackrel{\cdot \cdot }{a}}a=-\frac{\kappa ^2}6\left( \left( \rho
+\rho _{\text{de}}\right) +3\left( p+p_{\text{de}}\right) \right) ,
\end{equation}
where

\begin{equation}
\rho _{\text{de}}=\frac{18\alpha H^2\left( 6H^2-144\alpha H^4+\kappa
^2p\right) }{\kappa ^2\left( 1-30\alpha H^2\right) },
\end{equation}
\begin{equation}
p_{\text{de}}=\frac{6\alpha H^2\left( 6H^2+5\kappa ^2p\right) }{\kappa
^2\left( 1-30\alpha H^2\right) },
\end{equation}
are the density and the pressure of the ''dark energy'' $\alpha \Gamma ^2$.
The state equation of the ''dark energy'' is then
\begin{equation}
w_{\text{de}}=\frac{p_{\text{de}}}{\rho _{\text{de}}}=\frac 13\frac{%
6H^2+5\kappa ^2p}{6H^2-144\alpha H^4+\kappa ^2p}.
\end{equation}
The equations (8) and (9) yield
\begin{equation}
\stackrel{\cdot }{H}=-\frac{\kappa ^2}{2\left( 1-48\alpha H^2\right) }\left(
\rho +p\right) .
\end{equation}
In the vacuum we have a de Sitter solution
\begin{equation}
\stackrel{\cdot }{H}=0.
\end{equation}
For dust matter
\[
p=0
\]
the equations (8), (9), (11), (12) and (13) become
\begin{equation}
3H^2=108\alpha H^4\frac{1-24\alpha H^2}{1-30\alpha H^2}+\kappa ^2\rho ,
\end{equation}
\begin{equation}
\stackrel{\cdot }{H}=-\frac{3H^2\left( 1-18\alpha H^2\right) }{2\left(
1-30\alpha H^2\right) },
\end{equation}
\begin{equation}
\rho _{\text{de}}=\frac{108\alpha H^4\left( 1-24\alpha H^2\right) }{\kappa
^2\left( 1-30\alpha H^2\right) },
\end{equation}
\begin{equation}
p_{\text{de}}=\frac{36\alpha H^4}{\kappa ^2\left( 1-30\alpha H^2\right) },
\end{equation}
and
\begin{equation}
w_{\text{de}}=\frac 1{3\left( 1-24\alpha H^2\right) }.
\end{equation}
The evolution of $w_{\text{de}}$ is illustrated in Fig. 1.

\begin{figure}[ht]
  \includegraphics[width=6cm]{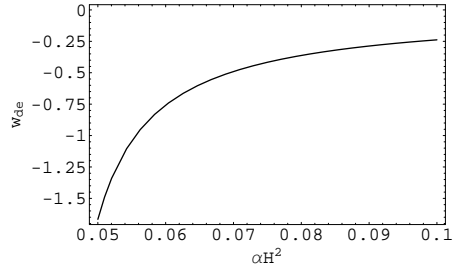}\\
  \caption{The evolution of $w_{de}$.}\label{wde}
\end{figure}

The equations (11), (12), and (13) indicate that the property of the ''dark
energy'' depends on the pressure $p$ of the ordinary matter. In dust matter
or vacuum, however, the property of the ''dark energy'' depends only on the
spacetime geometry as indicated by the equations (18), (19), and (20).

Letting
\[
w_{\text{de}}=-1
\]
and
\[
H_0=74km/sec/Mpc\simeq 2.4\times 10^{-18}sec^{-1}
\]
one can compute
\begin{equation}
\alpha =\frac 1{18\left( 74\right) ^2}=1.\,0145\times 10^{-5}\left(
km/sec/Mpc\right) ^{-2}=9.\,6451\times 10^{33}sec^2.
\end{equation}
Then letting
\begin{eqnarray}
w_{de}=-\frac{1}{3}\nonumber
\end{eqnarray}
we obtain
\begin{equation}
H=90.\,631km/sec/Mpc.
\end{equation}
Observations of type Ia supernovae at moderately large redshifts ($z\sim 0.5$
to $1$) have led to the conclusion that the Hubble expansion of the universe
is accelerating \cite{1}. This is consistent also with microwave background
measurements \cite{7}. According to the result of \cite{8}, $%
H=90.\,631km/sec/Mpc$ corresponds to
\[
z\sim 0.88,
\]
which is consistent with the observations.

Using the formula
\begin{equation}
H=\frac{\stackrel{\cdot }{a}}a=-\frac 1{1+z}\stackrel{\cdot }{z},
\end{equation}
we have
\[
\stackrel{\cdot }{H}=\frac{dH}{dt}=\frac{dH}{dz}\stackrel{\cdot }{z}%
=-H\left( 1+z\right) \frac{dH}{dz}.
\]
Then the equation (17) leads to
\[
-H\left( 1+z\right) \frac{dH}{dz}=-\frac{3H^2\left( 1-18\alpha H^2\right) }{%
2\left( 1-30\alpha H^2\right) },
\]
and can be integrated
\begin{equation}
^{}z=\left( \frac H{H_0}\right) ^{2/3}\left( \frac{18\alpha H^2-1}{18\alpha
H_0^2-1}\right) ^{2/9}-1.
\end{equation}
The function $z=z\left( H\right) $ is illustrated in Fig. 2 which is roughly
consistent with the observations \cite{8}.

The equation (8) indicates that during the evolution of the universe $H^2$
decreases owing to decreasing of the matter density $\rho $. This makes $w_{%
\text{de}}$ descend during the evolution of the universe. The evolution of
the function $w_{\text{de}}=w_{\text{de}}\left( \alpha H^2\right) $ given by
(20) is illustrated in Fig. 1. According to (20) when $H^2=\frac 1{12\alpha }
$, $w_{\text{de}}=$ $-\frac 13$, $\alpha \Gamma ^2$ changes from ''visible''
to dark as indicated by (10). If $H^2>\frac 1{12\alpha }$, it decelerates
the expansion, if $H^2<\frac 1{12\alpha }$, it accelerates the expansion.
When $H^2=\frac 1{18\alpha }$, $w_{\text{de}}$ crosses the phantom divide
line $-1$. In other words, the expansion of the universe naturally includes
a decelerating and an accelerating phase. $\alpha $ given by (21) can be
seen as a new constant describing the evolution of the universe.

\begin{figure}[ht]
  \includegraphics[width=6cm]{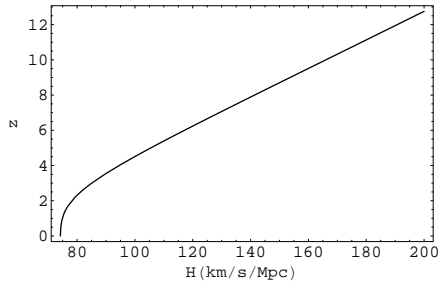}\\
  \caption{The function of $z=z(H)$.}\label{z-H}
\end{figure}


\begin{thebibliography}{*}

\bibitem{1}  Supernova Search Team, A. G. Riess et al., Astron. J. 116, 1009
(1998), astro-ph/9805201; Supernova Cosmology Project, S. Perlmutter et al.,
Astrophys. J. 517, 565 (1999), astro-ph/9812133; G. Hinshaw et al.,
Astrophys. J. Suppl. 148, 135 (2003), astro-ph/0302217; J. L. Tonry et. al.,
Astrophys. J. 594, 1 (2003).

\bibitem{2}  N. Straumann, astro-ph/0203330; E. V. Linder, arXiv:1009.1411
[astro-ph.CO]; K. Bamba, S. Capozziello, S. Nojiri and S. D. Odintsov,
arXiv:1205.3421 [gr-qc].

\bibitem{3}  T. P. Sotiriou and V. Faraoni, {\it Rev. Mod. Phys.} {\bf 82},
541 (2010); A. De Felice and S. Tsujikawa {\it Living Rev. Rel}. 13, 3
(2010); S. Capozziello and M. De Laurentis, {\it Phys. Rep}. {\bf 509}, 167
(2011); T. Clifton, P.G. Ferreira, A. Padilla and C. Skordis, {\it Phys. Rep}%
. {\bf 513}, 1 (2012).

\bibitem{4}  G. R. Bengochea and R. Ferraro, Phys. Rev. D79, 124019 (2009);
E. V. Linder, Phys. Rev. D81, 127301 (2010); P. Wu and H. Yu, Phys. Lett. B
692, 176 (2010); P. Wu and H. Yu, Phys. Lett. B 693, 415 (2010); P. Wu and
H. Yu, Eur. Phys. J. C 71, 1552 (2011); K. Bamba, C. Q. Geng and C. C. Lee,
JCAP, 08, 021 (2010); S. H. Chen, J. B. Dent, S. Dutta and E. N. Saridakis
(2010), Phys. Rev. D 83, 023508 (2011); G. R. Bengochea, Phys. Lett. B 695,
405 (2011); R. J. Yang, Europhys. Lett. 93, 60001 (2011); R. Zheng and Q.
-G. Huang (2010), J. Cosmol. Astropart. Phys. 03 (2011) 002; K. Bamba, C. Q.
Geng, C. C. Lee and L. -W. Luo, J. Cosmol. Astropart. Phys. 01 (2011) 021.

\bibitem{5}  B. Li, T. P. Sotiriou and J. D. Barrow, Phys. Rev. D 83, 064035
(2011); T. P. Sotiriou, B. Li and J. D. Barrow, Phys. Rev. D 83, 104030
(2011); B. Li, T. P. Sotiriou, and J. D. Barrow, Phys. Rev. D 83, 104017
(2011).

\bibitem{6}  T.T. Burwick, A.H. Chamseddine, K.Meissner, Phys.Lett. B284, 11
(1992), [arXiv:hep-th/9204015]; A. H. Chamseddine, Phys.Lett. B557, 247
(2003) [arXiv:hep-th/0301014]; B. M. Gripaios, JHEP0410, 069 (2004)
[arXiv:hep-th/0408127]; O. Bertolami, J. Paramos, Phys.Rev. D72, 044001
(2005), [arXiv:hep-th/0504215]; G. Helesfai, Class.Quant.Grav.25, 235010,
(2008), [arXiv:0806.3356];{\bf \ }S. S. Gubser, Fabio D. Rocha,
Phys.Rev.Lett.102, 061601 (2009), [arXiv:0807.1737]; R.Potting,
J.Phys.Conf.Ser.171, 012041 (2009), [arXiv:0904.0364]; S. Nojiri, S. D.
Odintsov, Phys.Rev.D83, 023001 (2011), [arXiv:1007.4856]; D. K. Wise, J.
Phys.: Conf. Ser. 360, 012017 (2012), [arXiv:1112.2390]; K. Krasnov,
arXiv:1112.5097; B. E. Meierovich, Phys.Rev.D79, 104027 (2009),
[arXiv:0910.1777]; E. T. Tomboulis, Phys. Rev. D 84, 084018 (2011),
[arXiv:1105.5848].

\bibitem{7}  C. L. Bennett et. al, WMAP team, Astrophys. J. 583, 1 (2003);
e-print arXiv: 1001.4758 (2010).

\bibitem{8}  J. Simon et al, 2005 Phys. Rev. D 71, 123001
[astro-ph/0412269]; A. G. Riess et al., Astrophys.J. {\bf 699}, 539 (2009)
[arXiv:0905.0695]; D. Stern, R. Jimenez, L. Verde, M. Kamionkowski and S. A.
Stanford, [arXiv:astro-ph/0907.3149].


\end{thebibliography}
\end{document}